\documentclass[%
 reprint,
 amsmath,amssymb,
 aps,
]{revtex4-1}

\usepackage{graphicx}
\usepackage{dcolumn}
\usepackage{bm}
\usepackage{color}

\usepackage{amsmath,amsfonts,amsbsy,amssymb}
\usepackage{indentfirst}
\usepackage{mathtools}

  \makeatletter
    \renewcommand{\theequation}{%
    \arabic{section}.\arabic{equation}}
    \@addtoreset{equation}{section}
  \makeatother

\newcommand{\fsl}[1]{\ensuremath{\mathrlap{\not{\phantom{#1}}}#1}}

\renewcommand{\BibitemShut}[1]{}

\newcommand{\nn}{\nonumber}

\def\be{\begin{equation}}
\def\ee{\end{equation}}
\def\bse{\begin{subequations}}
\def\ese{\end{subequations}}
\def\bal{\begin{align}}
\def\ealn{\end{align}}

\begin{document}

\preprint{APS/123-QED}

\title{Formulation of the Relativistic  Quantum Hall Effect 
 and ``Parity Anomaly''} 

\author{Kouki Yonaga$^1$, Kazuki Hasebe$^2$, and Naokazu Shibata$^1$}
\affiliation{%
$^1$Department of Physics, Tohoku University, Sendai, 980-8578, Japan  \\
$^2$Sendai National College of Technology, Ayashi, Sendai, 989-3128, Japan 
}%

\date{\today}

\begin{abstract}
We present a relativistic formulation of the quantum Hall effect on  {Haldane sphere}.  An explicit form of the pseudopotential is derived for the  relativistic quantum Hall effect with/without mass term. 
We clarify  particular features of the relativistic quantum Hall states with the use of the exact diagonalization study of the pseudopotential Hamiltonian. 
Physical effects of the mass term to the relativistic quantum Hall states are investigated in detail.    
The mass term acts as an interpolating parameter between the relativistic and non-relativistic quantum Hall effects.  
{
It is pointed out that the mass term unevenly affects the many-body
physics of the positive and negative Landau levels as a manifestation of
the ``parity anomaly''. In particular, we explicitly demonstrate the instability of the
Laughlin state of the positive first relativistic Landau level
with the reduction of the charge gap.}

\end{abstract}

\maketitle

\section{\label{sec:introduction}Introduction}

Dirac matter has attracted  considerable attention in condensed matter physics   
for its novel properties and recent experimental realizations in solid 
materials. 
In contrast to normal single-particle excitations in solids, 
Dirac particles exhibit  linear dispersion in a low energy region and
continuously vanishing density of states at the charge-neutral point \cite{Wallace-1947}.
These features are actually realized in graphene \cite{Zhou-et-al-2006, Bostwick-et-al-2007} and on topological insulator surface \cite{Zhang-et-al-2009}.  
Besides, in the presence of a magnetic field,   
the relativistic quantum Hall effect was observed in  graphene \cite{Zhang-et-al-2005, Novoselov-et-al-2005, Bolotin-et-al-2011} and also on topological insulator surface recently \cite{Yoshimi-et-al-2015, Xue-et-al-2014}.

\begin{figure}[t]
 \centering
 \includegraphics[width=6.5cm,clip]{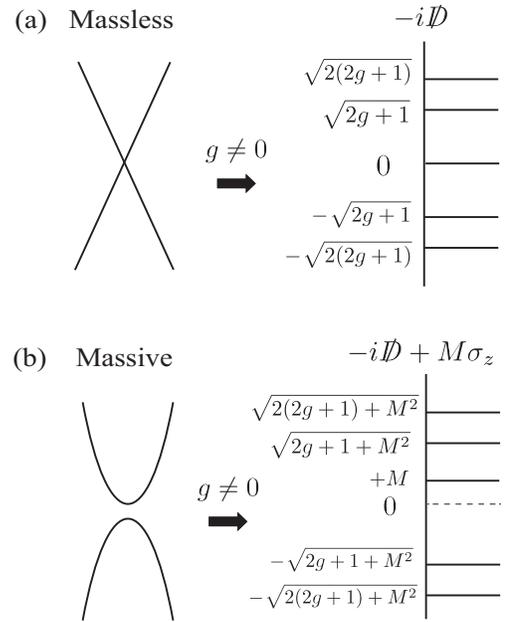}
 \caption{Schematic of the energy spectrum and Landau level. (a) and (b) show the massless and the massive cases ($M>0$), respectively. ($g$ represents a monopole charge.)  
The right figure of (b) shows asymmetry between the positive and negative energy levels due to  the absence of $-M$.  (In general, there exists the energy level  $E=+ \text{sgn}(g\cdot M)~|M| $ while not $E=-\text{sgn}(g\cdot M) ~|M| $.)  
The original reflection  symmetry of the energy levels with respect to the zero-energy is broken due to   the mass term.  }
 \label{fig:spectrum}
\end{figure}

One of the most intriguing features of the relativistic quantum Hall effect is the effect of  mass term; in the non-relativistic quantum Hall effect, the mass parameter just tunes  the Landau level spacing, while in the  relativistic quantum Hall effect the mass term is concerned with interesting physics such as the  semi-metal to insulator transition and the time reversal symmetry breaking of the topological insulators \cite{QiHZ2008}.  In experiments,     
 disorder and interaction with a substrate in the atomic layer of graphene cause the asymmetry in the two sublattices of a  
honeycomb structure to induce a mass term  
 \cite{Zhou-et-al-2007, Pereira-et-al-2008}, and  magnetic doping in topological insulators  yields a massive gap of the surface Dirac cone \cite{Chen-et-al-2010}.  Interestingly, in the presence of an external magnetic field, the mass term  brings the physics associated with the ``parity anomaly''    
  \footnote{The parity anomaly is usually referred to as the parity breaking Chern-Simons term  induced by the quantum effect. In the present paper the ``parity anomaly'' is referred to as  asymmetry of the positive and negative Landau levels due to  the (parity breaking) mass term in (2+1)D.   
}; in the absence of a magnetic field,  the mass term does not  change the equivalence between  the positive and negative energy levels \footnote{The mass term does not necessarily violate the reflection symmetry of the  Dirac operator spectrum  with respect to the zero energy. For instance in the case of the free Dirac operator,   the mass deformation breaks the chiral symmetry,  but the spectrum still preserves the reflection symmetry.  The chiral symmetry is a sufficient condition of the reflection symmetry of the spectrum but not a necessary condition.}, while in the presence ot a magnetic field,  asymmetry occurs between the positive  and negative energy levels depending on the sign of the mass parameter \cite{Haldane1988} (see Fig.\ref{fig:spectrum}).

In this paper, we establish a relativistic formulation of the quantum Hall effect on a two-sphere and perform a first investigation of the ``parity anomaly'' in the context of the relativistic $\it{many}$-$\it{body}$ physics. 
For concrete calculations, the spherical geometry called {Haldane sphere}  \cite{Haldane1983} 
is adopted. 
Instead of using the approximate pseudopotential of the infinite disk 
geometry in the previous study \cite{Shibata-Nomura-2009}, we construct an {\it exact} form of the pseudopotential based on the relativistic Landau model recently analyzed by one of the authors \cite{Hasebe-2015}. 
Previous numerical studies on fractional quantum Hall states in graphene show 
the existence of a Laughlin state at $\nu=1/3$ even in the $n=1$ Landau level \cite{Toke-et-al-2006, Apalkov-2006}
where the charge excitation gap is larger than that of the $n=0$ Landau level.
This stability of the Laughlin state in the $n=1$ Landau level
is a unique feature of the linear dispersion of the Dirac equation.
We study how this stability of the Laughlin states changes with 
 increase of  mass.  
Based on the exact diagonalization, we numerically obtain a many-body ground state  
of the relativistic pseudopotential Hamiltonian, and analyze the mass effect to the  Laughlin state at $\nu = {1}/{3}$
in the $n=1$ relativistic Landau level.

\section{Relativistic Landau problem on a sphere}{\label{sec:DiracEigenvalue}}

In this section, we give a brief review  of the relativistic Landau problem on the Haldane's sphere  \cite{Hasebe-2015} and discuss its mass deformation. 
The monopole gauge field is given by \cite{Wu-Yang-1976,Wu-Yang-1977}
\be
A=-g\cos\theta ~d\phi, 
\label{Schwingermonopolegauge}
\ee
where $g$ denotes the monopole charge. (In the following, we assume that $g$ is positive for simplicity.) 
{The Dirac operator on the Haldane sphere can be represented as $-i\fsl{{D}}=-ie_m^{~~\mu}\gamma^m(\partial_{\mu}+i\omega_{\mu}-iA_{\mu})$ where 
$e_{m}^{~~\mu}$ ($m=1,2$, $\mu=\theta, \phi$) denote the zweibein of two-sphere whose non-zero components are  $e_1^{~~\theta}=1/R$, $e_2^{~~\phi}={1}/(R\sin\theta)$ ($R$ is the radius of the Haldane sphere) and $\omega_{\mu}$ stands for the spin connection.   
When we adopt the 2D gamma matrices as $(\gamma^1, \gamma^2) =(\sigma_x, \sigma_y)$, the spin connection is expressed as $\omega_{\theta}=0, \omega_{\phi}=\frac{1}{2}\sigma_z$, and then the Dirac operator takes the form of }
\begin{align}
-i\fsl{{D}}&=-i\frac{1}{R}\sigma_x \partial_{\theta}-i\frac{1}{R\sin\theta}\sigma_y (\partial_{\phi}+i(g-\frac{1}{2}\sigma_z)\cos\theta)\nonumber\\
&=-i\frac{1}{R}\sigma_x (\partial_{\theta}+\frac{1}{2}\cot\theta)-i\sigma_y \frac{1}{R\sin\theta}(\partial_{\phi}+ig\cos\theta),  
\label{explicidiragaugesch}
\end{align}
or 
\begin{equation}
-i\fsl{{D}}
=\frac{1}{R}\begin{pmatrix}
0 & -i\eth_-^{(g+\frac{1}{2})} \\
-i\eth_+^{(g-\frac{1}{2})} & 0 
\end{pmatrix}.   \label{diracopfrometh}
\end{equation}
Here 
 $\eth_\pm^{(g+\frac{1}{2})}$ are  the  edth operators \cite{Newman-Penrose-1966}: 
\be
\eth_{\pm}^{(g)} 
={\partial_\theta}\mp ig\cot\theta \pm i\frac{1}{\sin\theta}{\partial_{\phi}}.
\label{expliladdercharge}
\ee
For graphene, two components of the Dirac spinor indicate  
 sublattice degrees of freedom, while for a topological insulator 
 the real spin degrees of freedom of surface electron.  
%
The eigenvalues of the Dirac operator (\ref{explicidiragaugesch}) are derived as 
\be
\pm \lambda_n =\pm \frac{1}{R}\sqrt{n(n+2g)}, ~~~~(n=0, 1, 2, \cdots)
\label{eigenvaluesofdirac}
\ee
where $n$ corresponds  to the relativistic Landau level index. 
Notice that the spectrum (\ref{eigenvaluesofdirac}) exhibits the reflection symmetry with respect to the zero energy.  
Each Landau level $\pm \lambda_n$ accommodates the following degeneracy 
\be
d_{ \lambda_n}=d_{-\lambda_n}=2(g+n). 
\label{degelandaulevel}
\ee
The degenerate eigenstates of the Landau level are   
\begin{subequations}
\begin{align}
&n=1,2,\cdots:\nn\\
&~~\psi_{\pm \lambda_n, m}^g(\theta, \phi) 
= \frac{1}{\sqrt{2}}\begin{pmatrix}
{Y}^{g-\frac{1}{2}}_{j=(g-\frac{1}{2})+n ,m}(\theta, \phi) \\
\mp i{Y}^{g+\frac{1}{2}}_{j=(g+\frac{1}{2})+(n-1) ,m} (\theta, \phi)
\end{pmatrix}, 
\label{eigenstateofdiracope}\\
&n=0~:~\psi_{\lambda_0=0, m}^g(\theta, \phi) 
= \begin{pmatrix}
{Y}^{g-\frac{1}{2}}_{j=g-\frac{1}{2} , m}(\theta, \phi) \\
0
\end{pmatrix}, 
\label{eigenstateofdiracopezero}
\end{align}\label{eigenstateofdiracop}
\end{subequations}
with $m=-g+\frac{1}{2}+n, -g+\frac{3}{2}+n, \cdots, g-\frac{1}{2}+n$.    ${Y}^{g}_{j ,m}$ denote the monopole harmonics \cite{Wu-Yang-1976}: 
\begin{align}
&{Y}^g_{~l,m}(\theta, \phi)
=2^m \sqrt{\frac{(2l+1)(l-m)!(l+m)!}{4\pi (l-g )!(l+g)!}} \nn\\
&\times (1-x)^{-\frac{m+g}{2}} (1+x)^{-\frac{m-g}{2}}P_{l+m}^{(-m-g, -m+g)} (x) \cdot e^{im\phi},   
\label{monopoleharmonicsjacobischwinger}
\end{align}
where $x=\cos\theta$ and $P_{n}^{(\alpha, \beta)} (x)$ stand for the Jacobi polynomials.  

{The magnetic field does not affect the spectrum symmetry between the positive and negative Landau levels (\ref{eigenvaluesofdirac}), but acts unevenly on  the upper and lower components of the eigenstates, which can most apparently be seen from the absence of the lower component of the zero-mode (\ref{eigenstateofdiracopezero}).}    
Also notice that the components of $\psi^{g}_{\pm\lambda _n, m}$ (\ref{eigenstateofdiracope}) consist of the monopole harmonics in different non-relativistic Landau levels, $n$ and $n-1$, and   carry the same $SU(2)$ index 
\be
j=g-\frac{1}{2}+n, 
\label{jandng}
\ee
which implies that $\psi_{\pm\lambda_n, m}^g$ itself transforms as the $SU(2)$ irreducible representation. 
Such  $SU(2)$  angular momentum operators are given by   
\be
J_i=-i\epsilon_{ijk}x_j(\partial_k -i\mathcal{A}_k) -(g-\frac{1}{2}\sigma_z)\frac{x_i}{r},  
\label{su2angularmom}
\ee
with 
\be
\mathcal{A}=\mathcal{A}_i dx_i=-(g-\frac{1}{2}\sigma_z)\cos\theta d\phi. 
\ee
$J_i$ (\ref{su2angularmom}) is {formally equivalent} to the total angular momentum of the non-relativistic charge-monopole system  with the replacement of the monopole charge $g$ to a matrix value, $g-\frac{1}{2}\sigma_z$. 
The Dirac operator is a singlet under the $SU(2)$ transformation generated by $J_i$, 
\be
[J_i, -i\fsl{D}]=0, 
\ee
and so there exist the simultaneous eigenstates (\ref{eigenstateofdiracop})  of the Dirac operator and the $SU(2)$ Casimir. Each relativistic Landau level thus accommodates the $SU(2)$ degeneracy (\ref{degelandaulevel}), 
$2j+1=2(g+n)$.

The Dirac operator also respects the chiral symmetry, 
\be
\{-i\fsl{D}, \sigma_z\}=0, 
\ee
and the spectrum of the Dirac operator is symmetric with respect to the zero eigenvalue (\ref{eigenvaluesofdirac}). 
 The non-zero Landau level eigenstates of the same eigenvalue magnitude   (\ref{eigenstateofdiracope}) are related by the chiral transformation: 
\be
\psi_{\pm \lambda_n, m}^{g}=\sigma_z \psi_{\mp \lambda_n, m}^{g}. 
\ee

\subsection{Mass deformation}\label{subsec:massdeformation}

\begin{figure}[t]
 \centering
 \includegraphics[width=8cm]{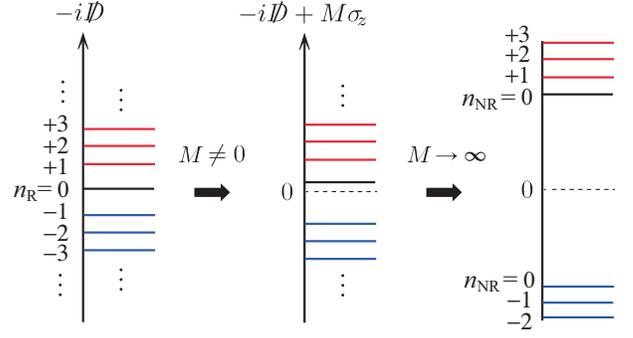}
 \caption{(Color online) 
 The change of the Dirac operator spectrum under the mass deformation. 
 $n_{\rm R}$ corresponds to the relativistic Landau level by $\text{sgn}(n_{\rm R}) \cdot \Lambda_{n=|n_{\rm R}|}$, and 
 $n_{\rm N\!R}$ to the non-relativistic Landau level by $\text{sgn}(n_{\rm N\!R}) \cdot E_{n=|n_{\rm N\!R}|}$. 
The mass deformation brings the asymmetry of the spectrum due to the presence of $n_{\rm R}=0$ Landau level $\Lambda_{n=0}=+M$ (the black solid line in the middle figure).   
In the non-relativistic limit (the right figure), the spectrum yields the non-relativistic Landau level spectrum.  Notice that each of the negative relativistic Landau levels decreases (the absolute value of) its Landau level index by one in the non-relativistic Landau level as  $|n_{\rm N\!R}|=|n_{\rm R}|-1$.   }
 \label{fig:Massdef}
\end{figure}

We apply a mass deformation to the Dirac operator:  
\be
-i\fsl{{D}}+\sigma_z M =\frac{1}{R}\begin{pmatrix}
R M & -i\eth_-^{(g+\frac{1}{2})} \\
 -i\eth_+^{(g-\frac{1}{2})}    & -R M 
\end{pmatrix}. 
\label{mass2dhamildirac}
\ee
The mass deformation does not break the $SU(2)$ rotational symmetry of the system 
\be
[\sigma_z M, {J}_i]=0, 
\ee
but break the chiral symmetry, 
$\{\sigma_z M, \sigma_z\}=2M \neq 0.$ 
{The mass deformation is expected to bring asymmetry between  the positive and negative spectrum. }  
The square of the mass deformed Dirac operator yields 
\be
(-i\fsl{{D}}+\sigma_z M)^2=(-i\fsl{{D}})^2+M^2,  
\ee
and the eigenvalues of the mass deformed Dirac operator are obtained as  
\begin{subequations}
\begin{align}
&n= 1, 2, \cdots:\pm \Lambda_n 
=\pm \frac{1}{R}\sqrt{n(n+2g)+(MR)^2},   \label{nonzeromassivelandausp}\\
&n=0~~~~~~~~:~\Lambda_0=+M.  \label{zeromassivelandausp}
\end{align}\label{eigenvaluesrelat}
\end{subequations}
Note that for $n\ge 1$ the spectra (\ref{nonzeromassivelandausp}) still respect the reflection symmetry  with respect to the zero energy, while $n=0$ (\ref{zeromassivelandausp}) does not have its counterpart of the negative energy $-M$ [Fig.\ref{fig:Massdef}]. This asymmetry is a manifestation of the ``parity anomaly'' {as mentioned in Introduction}.   
The corresponding eigenstates for (\ref{eigenvaluesrelat}) are respectively given by 
\begin{subequations}
\begin{align}
&n=1,2,\cdots:\nn\\
&\psi^{g}_{\pm \Lambda_n, m}=\sqrt{\frac{\Lambda_n+\lambda_n}{2\Lambda_n}} ~(\psi^{g}_{\pm \lambda_n, m} \pm \frac{M}{\Lambda_n+\lambda_n} \psi^{g}_{\mp \lambda_n, m}), \label{nonzeromassivelandauspfunc}\\
&n=0~~~~:~\psi^{g}_{\Lambda_0=M, m}= \psi_{\lambda_0=0, m}^{g}, 
\label{eigenvaluemassivedirac}
\end{align}
\end{subequations}
where $m=-n-g+\frac{1}{2},-n-g+\frac{3}{2}, \cdots, n+g-\frac{1}{2}$. 
One may find that $\psi^{g}_{\Lambda_0=M, m}$ (\ref{eigenvaluemassivedirac}) are simply the zero-modes of the massless Dirac operator, while  $\psi^{g}_{+ \Lambda_n, m}$ and $\psi^{g}_{-\Lambda_n, m}$ (\ref{nonzeromassivelandauspfunc}) are  linear combinations of the chirality partners of non-zero Landau levels, 
$\psi_{+\lambda_n, m}^g$ and $\psi_{-\lambda_n, m}^g$,  and reduced to them   
  in  $M\rightarrow 0$. 
The mass deformation is inert to the  Landau level degeneracy; 
 each of the Landau levels $\pm\Lambda_n$ has the same degeneracy    $d_{\pm\Lambda_n}=2(g+n)$, since the $SU(2)$ symmetry is kept exact even under   the mass deformation.

Meanwhile in $M\rightarrow \infty$ which we call the non-relativistic limit,  the positive Landau level spectra $(\ref{nonzeromassivelandausp})$ are reduced to 
\be
+ \Lambda_n~\simeq~ M+\frac{1}{2MR^2}n(n+2g)+O(\frac{1}{M^3}), 
\label{posieigenvalunon}
\ee
and the eigenstates (\ref{nonzeromassivelandauspfunc}) are  
\be
\psi^{g}_{+ \Lambda_n, m}~\simeq ~\frac{1}{\sqrt{2}}(\psi_{\lambda_n, m}^{g}+\psi_{-\lambda_n, m}^{g} )=
 \begin{pmatrix}
Y^{g-\frac{1}{2}}_{j=g-\frac{1}{2}+n, m} \\
0
\end{pmatrix}. \label{eigenreduceposi}
\ee
Notice that the next leading order term on the right-hand side of (\ref{posieigenvalunon}) is equal to  the non-relativistic Landau levels \cite{Haldane1983}:   
\be
E_n=\frac{1}{2MR^2} (n(n+2g+1)+g), 
\label{nonrelalandau}
\ee
with replacement from $(n, g)$ to $(n, g-\frac{1}{2})$ up to a unimportant constant  and for  the eigenstates (\ref{eigenreduceposi}) also. 
Thus in the non-relativistic limit,  $n$th  positive relativistic Landau level is reduced to  $n$th non-relativistic Landau level  with monopole charge $g-\frac{1}{2}$. 
Similarly for the negative relativistic Landau level, we have 
\be
- \Lambda_n~\simeq~ -M-\frac{1}{2MR^2}n(n+2g)+O(\frac{1}{M^3}), 
\label{negaeigenvalunon}
\ee
and 
\be
\psi^{g}_{- \Lambda_n, m}~\simeq ~\frac{1}{\sqrt{2}}(\psi_{-\lambda_n, m}^{g}-\psi_{\lambda_n, m}^{g} )=
 i\begin{pmatrix} 
0\\
Y^{g+\frac{1}{2}}_{j=g+\frac{1}{2}+(n-1), m}  
\end{pmatrix}.   \label{eigenreducenega}
\ee
The next leading order term of the opposite sign on the right-hand side of (\ref{negaeigenvalunon}) is equal to  the non-relativistic Landau level (\ref{nonrelalandau}) 
with replacement from $(n, g)$ to $(n-1, g+\frac{1}{2})$  and for the eigenstates (\ref{eigenreducenega}) also.  
Thus in the non-relativistic limit,  up to a constant, (the absolute value of) the  $n$th negative relativistic Landau level reproduces  $(n-1)$th non-relativistic Landau level  with monopole charge $g+\frac{1}{2}$. 
 
To summarize, the mass parameter interpolates the $n$th positive/negative relativistic Landau level physics ($M\rightarrow 0$) and the 
$n/n-1$th non-relativistic Landau level physics ($M \rightarrow \infty$).  It is important to note that the positive and negative relativistic Landau levels approach to the different non-relativistic Landau levels in the $M\rightarrow\infty$ limit. 
We will revisit this in the context of many-body physics 
in Sec.\ref{seq:results}.

\section{Relativistic Pseudopotential \label{sec:pp}}

We next construct a relativistic pseudopotential on the  Haldane sphere.  
Neglecting Landau level mixing, the projection Hamiltonian onto the $n$th Landau level is given by
\begin{equation}
H = \sum_{p<q} \sum_{J} V^{n}_{J} P_J(p,q) 
\label{defprojectionHamiltonian}
\end{equation}
where $V^{n}_{J}$ is Haldane's pseudopotential in $n$th Landau level and $P_J(p,q)$ projects onto states in which $p$th and $q$th particles have two-body angular momentum $J$.
$P_J(p,q)$ is given by
\begin{equation}
P_J(p,q) = \prod_{J'\neq J} \frac{ ( \bm{J}_{p} + \bm{J}_{q} )^{2} - J'(J'+1) }{ J(J+1)-J'(J'+1) } \label{projectionops}
\end{equation}
where $\bm{J_}{p}$ (\ref{su2angularmom}) represents the angular momentum operator for the $p$th particle and $J(J+1)$ is the eigenvalue of $( \bm{J}_{p}+\bm{J}_{q})^{2}$. 
Due to the Wigner-Eckart theorem, the pseudopotential can be expressed as 
\begin{equation}
V^{n}_{J}\delta_{J_{z},J'_{z}} = \langle\!\langle J,J_{z} | V | J,J'_{z} \rangle\!\rangle
\label{eq:VJ}
\end{equation} 
where $V$ is Coulomb interaction and $|J,J_{z} \rangle\!\rangle$ denotes a two-particle state with azimuthal angular momentum $J_{z}$.  

Let us begin with  the massless case. For notational brevity, we rewrite the relativistic eigenstates  (\ref{eigenstateofdiracope}) as the state vector $|j,m,g\rangle\!\rangle$:  
\begin{equation}
|j,m,g\rangle\!\rangle_{\pm\lambda_{n}} = \frac{1}{\sqrt[]{2}}( |j,m,g-\frac{1}{2}\rangle|\! \uparrow\rangle \mp i | j,m,g+\frac{1}{2} \rangle |\!\downarrow\rangle ) 
\label{eq:harmonic_state}
\end{equation}
where
\begin{eqnarray}
&&\langle \theta, \phi | j,m,g\pm \frac{1}{2} \rangle = Y^{g\pm \frac{1}{2}}_{j,m}(\theta,\phi), \nonumber \\
&&~~~~~|\!\uparrow\rangle =\begin{pmatrix}
1 \\
0
\end{pmatrix},~~~|\!\downarrow\rangle =\begin{pmatrix}
0 \\
1
\end{pmatrix}. 
\end{eqnarray}
The two-particle state $|J,J_{z} \rangle\!\rangle$ is given by 
\begin{equation}
|J,J_{z} \rangle\!\rangle_{\pm\lambda_{n}} = \sum_{m_{1},m_{2}} C^{J,J_{z}}_{jm_{1},jm_{2}} |j,m_{1},g \rangle\!\rangle_{\pm\lambda_{n}} \otimes |j,m_{2},g \rangle\!\rangle_{\pm\lambda_{n}}
\label{eq:two-state}
\end{equation}
where $C^{J,J_{z}}_{jm,jm'}$ represents Clebsch-Gordan coefficients: 
\be
C^{J,J_{z}}_{jm,jm'}=
(-1)^{J_z} \sqrt{2J+1}
\begin{pmatrix}
j & j& J \\
m & m' & -J_z 
\end{pmatrix}. \label{defCG3j} 
\ee
$(\cdots)$ denotes the Wigner 3j-symbol.
Substituting Eq.~(\ref{eq:harmonic_state}) to Eq.~(\ref{eq:two-state}), we have 
\begin{eqnarray}
|J,J_{z} \rangle\!\rangle_{\pm\lambda_{n}} &=& \frac{1}{2} \sum_{m_{1},m_{2}} C^{J,J_{z}}_{jm_{1},jm_{2}} \nonumber \\
                              &\times& [ | j,m_{1},g-\frac{1}{2};j,m_{2},g-\frac{1}{2} \rangle |\!\uparrow,\uparrow\rangle \nonumber \\
                              &\mp& i | j,m_{1},g-\frac{1}{2};j,m_{2},g+\frac{1}{2} \rangle |\!\uparrow,\downarrow\rangle  \nonumber \\
                              &\mp& i| j,m_{1},g+\frac{1}{2};j,m_{2},g-\frac{1}{2} \rangle |\!\downarrow,\uparrow \rangle \nonumber \\
                              &-& | j,m_{1},g+\frac{1}{2};j,m_{2},g+\frac{1}{2} \rangle |\!\downarrow,\downarrow \rangle ]. \label{explicjjz}
\end{eqnarray}
With (\ref{explicjjz}), the relativistic pseudopotential (\ref{eq:VJ}) is  evaluated as 
\begin{equation}
V^{\rm R}_{J} = \frac{1}{4} \sum_{ \alpha,\beta = g \pm \frac{1}{2} } V_{J}(j,\alpha,\beta) 
\label{relativepseudopo}
\end{equation}
where 
\begin{align}
V_{J}(j,\alpha,\beta) =&\frac{1}{R} (2j+1)^{2}(-1)^{\alpha+\beta+J} \nonumber\\
&\!\!\!\!\!\!\!\!\!\!\!\!\!\!\!\!\!\!\times \sum_{k=0}^{2j} \begin{Bmatrix}
J & j & j \\
k & j & j 
\end{Bmatrix} \begin{pmatrix}
j & k & j \\
-\alpha & 0 & \alpha
\end{pmatrix}   \begin{pmatrix}
j & k & j \\
-\beta & 0 & \beta
\end{pmatrix}. \label{comppseudopoterela}
\end{align}
$\{\cdots \}$ denotes the 6j symbol. {(See Appendix \ref{seq:drv_pp} for detail derivation of (\ref{comppseudopoterela}) and  definition of the 6j-symbol.)}   
Notice that the relativistic pseudopotential (\ref{relativepseudopo}) is the average of the four pseudopotentials $V_J(j, \alpha, \beta)$ with $(\alpha, \beta)=(g-\frac{1}{2}, g-\frac{1}{2})$,   $(g+\frac{1}{2}, g+\frac{1}{2})$,  $(g-\frac{1}{2}, g+\frac{1}{2})$  and  $(g+\frac{1}{2}, g-\frac{1}{2})$.   
The first two correspond to the pseudopotentials of $n$th and ($n-1$)th non-relativistic Landau levels respectively. (Remember $j=g-\frac{1}{2}+n=g+\frac{1}{2}+(n-1).$)  
The remaining two come from the cross term between $n$th and ($n-1$)th non-relativistic Landau levels, which are  unique pseudopotentials  in the relativistic case.  

Next we move to the massive case.
The relativistic eigenstate (\ref{nonzeromassivelandauspfunc}) can be rewritten as 
\begin{align}
|j,m,g,M\rangle\!\rangle_{\pm\Lambda_{n}}& =\frac{1}{2}\sqrt{\frac{\Lambda_n+\lambda_n}{\Lambda_n}} \nn\\
&\!\!\!\!\!\!\!\!\!\!\!\!\!\!\!\! \!\!\!\!\times
\biggl[ \biggl(1\pm \frac{M}{\Lambda_n+\lambda_n}\biggr)|j, m, g-\frac{1}{2} \rangle |\!\uparrow \rangle \nn\\
&\!\!\!\!\!\!\! \!\!\!\! \!\!\mp i\biggl(1 \mp \frac{M}{\Lambda_n+\lambda_n}\biggr)|j, m, g+\frac{1}{2} \rangle |\! \downarrow \rangle  \biggr],
\end{align}
and in a similar manner to the massless case, the  pseudopotentials are derived as  
\begin{align}
&V^{\rm R}_{J}(M)_{\pm \Lambda_{n}} =\frac{1}{ 4 } \biggl [ \biggl(1\pm \frac{M}{\Lambda_n}\biggr)^2 ~V_J(j, g-\frac{1}{2}, g-\frac{1}{2})\nn\\
&~+\biggl(\frac{\lambda_n}{\Lambda_n}\biggr)^2 ~(V_{J}(j,g-\frac{1}{2},g+\frac{1}{2}) +V_{J}(j,g+\frac{1}{2},g-\frac{1}{2}))\nn\\
&~+\biggl(1\mp \frac{M}{\Lambda_n}\biggr)^2 ~V_J(j, g+\frac{1}{2}, g+\frac{1}{2}) \biggr].  \label{eq:ppM}
\end{align}
{
It is easy to confirm that (\ref{eq:ppM})  is reduced to (\ref{relativepseudopo}) in the massless limit $M\rightarrow 0$. Superficially the mass gap might not seem to have something to do with the Coulomb interaction, but quite interestingly  the mass gap indeed alters the strength of the effective Coulomb  pseudopotential as found in (\ref{eq:ppM}).}   
Since  the $SU(2)$ generators are immune to the mass deformation, the mass deformation affects the projection Hamiltonian (\ref{defprojectionHamiltonian}) {not through the projection operators (\ref{projectionops}) but through the pseudopotentials  (\ref{eq:ppM}) only. } 

In the following discussions, we express the pseudopotentials as a function of the relative angular momentum $m=2j-J$: {The readers should notice that $V_m$ in the following sections indicates $V_{J=2j-m}$ used in this section.}

\begin{figure}[t]
 \centering
 \includegraphics[width=7.0cm,clip]{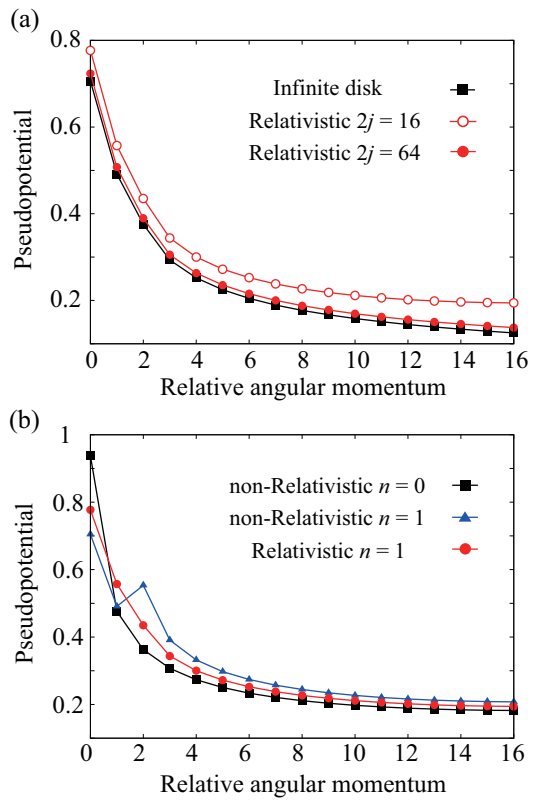}
 \caption{(Color online) (a) Total flux dependence of the pseudopotential for the relativistic massless particles (\ref{relativepseudopo}) in $n=1$. The open and filled circles show $V_{m}^{\rm R}$ with $2j=16$ and $64$, respectively. 
The solid squares represent the pseudopotential for the infinite disk geometry corresponding to  $2j=16$. (b) Pseudopotentials in several cases. The filled squares and triangles exhibit $V^{\rm nR}_{m}$ in $n=0$ and $n=1$. The solid circles represent the relativistic pseudopotential in $n=1$. {(The relativistic pseudopotential in $n=0$ is  the same as the non-relativistic pseudopotential in $n=0$.)} } 
 \label{fig:pp}
\end{figure}

\begin{figure}[t]
 \centering
 \includegraphics[width=8.5cm,clip]{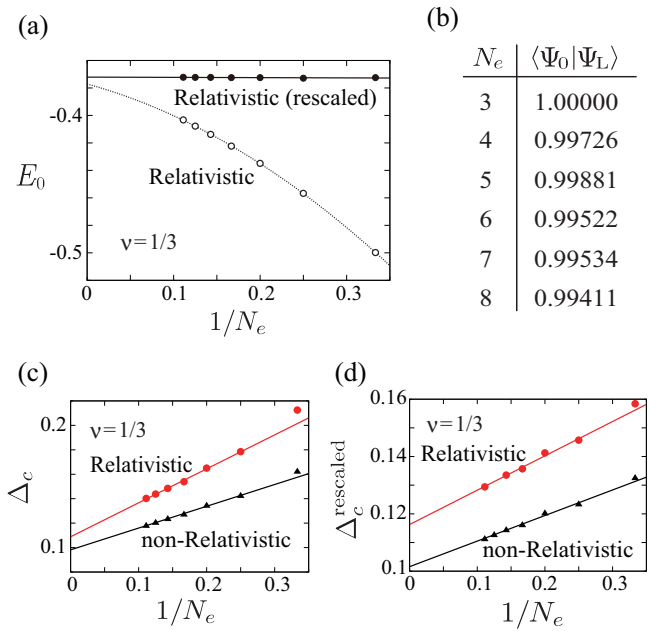}
 \caption{(Color online) (a) Ground state energies {at $\nu=1/3$ of relativistic $n=1$ Landau level} as a function $1/N_{e}$. The open and filled circles represent the lowest energies in  unit of $e^{2}/\epsilon l_{B}$ and $e^{2}/\epsilon l'_{B}$, respectively. The solid (dotted) line shows the linear (square) fitting for  each of the  energies. (b) Overlaps for each $N_{e}$
{between the ground state at $\nu=1/3$ of relativistic
$n=1$ Landau level and the Laughlin wave function}. (c) and (d) respectively show the charge gap and the rescaled one {at $\nu=1/3$ of relativistic $n=1$ (red circles) and non-relativistic $n=0$ (black triangles) Landau levels}. }
 \label{fig:energy}
\end{figure}

\section{ \label{seq:results} Numerical Results}

To investigate the physics of the mass deformation on many-body states, 
we perform an exact diagonalization study 
at $\nu={1}/{3}$. 
We focus on the single Dirac Hamiltonian, which describes the surface electrons of a 3D topological insulator or graphene with  full spin and valley polarization.  
In the following, we set $e^{2}/\epsilon l_{B}$ to unity.~($\epsilon$ is a dielectric constant.) 
It is well known that the ground state is described by the Laughlin state at $\nu={1}/{q}$ ($q$ is odd integer) \cite{Laughlin-1983}. 
In the spherical geometry, the Laughlin state is realized when the total flux ${N_{\Phi}}$ is given by \cite{Haldane1983,Haldane-Rezayi-1985}
\begin{equation}
{N_{\Phi}} = 2j = \nu^{-1}( N_{e} - 1 )
\end{equation}
where $N_{e}$ is the number of electrons in a partially filled Landau level.
We define the total energy as 
\begin{equation}
E({N_{\Phi}}) = E_{\rm C}({N_{\Phi}}) - \frac{N^{2}_{e}}{ 2R }
\end{equation}
where the first term represents the energy of Coulomb interaction and the second term means the effect of the neutralizing background and the self-energy of the background.

We also assess the energy gap for the creation of a quasiparticle or a quasihole as 
\begin{equation}
\Delta_{c}^{\pm} = E({N_{\Phi}}\pm1)-E({N_{\Phi}})
\end{equation}
where $+/-$ indicates to  quasihole/quasiparticle. 
{ A thermal excitation  is a neutral pair of a quasiparticle and a quasihole whose excitation gap is given by  
\be
\Delta_{c} = \Delta_{c}^{+} + \Delta_{c}^{-} 
\label{sumdeltas}
\ee
for infinite separation.}

\subsection{\label{seq:massless_result} Massless case}

We first investigate the massless relativistic case for comparison to the non-relativistic results obtained by Fano $\it{et~al}$. \cite{Fano-et-al-1986}.  
We perform numerical calculations only for $+\lambda_{n}$ since the pseudo-potential Hamiltonian of $-\lambda_{n}$ 
is equivalent to that of  $+\lambda_{n}$. 
Figure \ref{fig:pp} (a) shows the total flux dependence of the relativistic pseudopotential $V_{m}^{\rm R}$ in $n = 1$ and that of the pseudopotential $V^{\rm Disk}_{m}$ in infinite disk geometry \cite{Prange-Girvin-1987}: 
\begin{equation}
V^{\rm Disk}_{m} = \int \frac{dq}{2\pi} qV(q) [F_{n}(q)]^{2} L_{m}(q^{2})e^{-q^{2}}.
\end{equation}
Here $F_{n}(q)$ is the relativistic form factor \cite{Nomura-MacDonald-2006}
\begin{equation}
F_{n}(q) = \frac{1}{2} \left[ L_{n}(q^{2}/2) + L_{n-1}(q^{2}/2) \right]
\end{equation}
with Laguerre polynomials $L_{n}(x)$.

It is expected that $V^{\rm R}_{m}$ and $V_{m}^{\rm Disk}$ become equivalent in the thermodynamic limit ${g} \rightarrow \infty$ since the finite size effect of the sphere will vanish in { $R =\sqrt[]{g}l_{B}  \rightarrow \infty$ (with fixed $l_B=1/\sqrt{eB}$).}   
Indeed, with the increase of the  total flux {$2g$},   $V^{\rm R}_{m}$ apparently approaches   $V_{m}^{\rm Disk}$ in Fig.~\ref{fig:pp} (a). 
Thus we confirm the equivalence between the disk and spherical geometries in the thermodynamic limit. 

{The difference  between the relativistic and the non-relativistic pseudopotentials is expected to appear in the charge excitation energy of the Laughlin state at $\nu=1/3$. Since the charge gap of the Laughlin state strongly depends on $V_1-V_3$ [26], we focus on $V_1-V_3$. 
In Fig.~\ref{fig:pp}(b) we compare the $m$-dependence of the pseudopotentials between the relativistic $V_{m}^{\rm R}$ in $n = 1$ and the non-relativistic $V_{m}^{\rm nR}(= V_{J}(j,g,g)|_{j=n+g}) $ in $n=0$ and $1$. }
The value of $V_{1}-V_{3}$ of non-relativistic $n=1$ Landau level is the smallest, and hence its  Laughlin state is considered to be unstable {against large thermal and impurity effects in the energy unit of $e^2/\epsilon \ell_B$ in the experimental situation.} (Also notice that the pseudopotential shows a cusp at the relative angular momentum $m=2$.)  
Meanwhile,  the $n=1$ relativistic pseudopotential $V_{m}^{\rm R}$   shows a monotonic decay as a function of the relative angular momentum and  its $V^{\rm R}_{m=1}-V^{\rm R}_{m=3}$ is the largest.   Therefore, the Laughlin state in the $n=1$ relativistic level is expected to be more stable than those of the non-relativistic levels $n=0, 1$ \cite{Apalkov-2006,Shibata-Nomura-2009}.  

{To obtain the  properties of the bulk limit,  we need the extrapolation on the size of systems. We, however, sometimes have strong finite-size effects (for example, see the open circles in Fig.~\ref{fig:energy}(a)). Fortunately, with the rescaled magnetic length $l'_{B} = \sqrt{{N_{\Phi}}\nu/N_{e}}l_{B}=\sqrt{(N_e-1)/N_e} l_B$,  the ground state energy and the charge gap in units of $e^{2}/\epsilon l'_{B}$ give us good extrapolated values even if the number of electrons is small \cite{Morf-et-al-2002}. Indeed, the open circles in Fig.~\ref{fig:energy}(a) showing the lowest energies for each $N_{e}$ in the unit of $e^2/\epsilon \ell_B$ have large size dependence compared with that of the filled circles in Fig.~\ref{fig:energy}(a),  which correspond to the energies rescaled by $l'_{B}$.  The filled circles are scaled by a linear function to have  $E_{0}({{N_{e}} \rightarrow \infty}) \approx -0.3721$.} 

Figure \ref{fig:energy}(b) exhibits the overlaps $\langle \Psi_{0} | \Psi_{\rm L} \rangle$ between the numerical ground state $|\Psi_{0}\rangle$ 
{at $\nu=1/3$ of the relativistic $n=1$ Landau level} and the Laughlin wave function $|\Psi_{\rm L}\rangle$. The Laughlin state has a remarkably large overlap $\langle \Psi_{0} | \Psi_{\rm L} \rangle$ for each $N_{e}$ more than $99 \%$, which strongly suggests that the many-body groundstate of the relativistic $n=1$ Landau level is the Laughlin state.  

Figure \ref{fig:energy}(c) displays the charge gap $\Delta_{c}$ (\ref{sumdeltas}) {at $\nu=1/3$ of the relativistic $n=1$ and non-relativistic $n=0$ Landau levels}, and Fig.~\ref{fig:energy}(d)  exhibits the rescaled charge gap $\Delta^{\rm rescaled}_{c}$.  
The charge gap in the relativistic $n=1$ Landau level is larger than that in the non-relativistic $n=0$ Landau level.  
This is expected from the previous observation that the Laughlin state in the relativistic Landau level is more stable than in the non-relativistic Landau level.   
 From the rescaled charge gap, we obtain $\Delta_{c} \approx 0.1163$ {in the relativistic $n=1$ Landau level}. This value is in agreement with previous work using $V_{m}^{\rm Disk}$ in larger systems \cite{Shibata-Nomura-2009}.

\subsection{\label{seq:massive_result} Massive case}

\begin{figure}[t]
 \centering
 \includegraphics[width=7.0cm,clip]{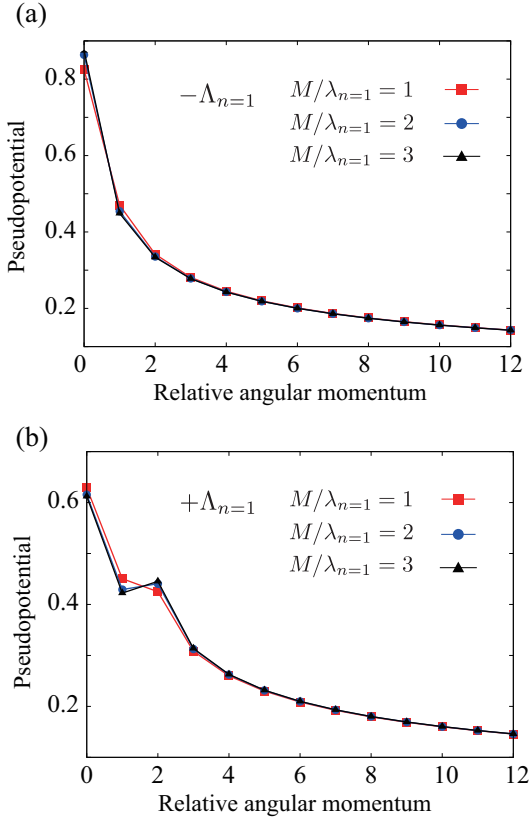}
 \caption{(Color online) Mass dependence of the pseudopotential with {$2j=64$}. (a) and (b) show the numerical results in 
{$-\Lambda_{n=1}$ and $+\Lambda_{n=1}$}, respectively. \\ }
 \label{fig:ppM}
\end{figure}

\begin{figure}[t]
 \centering
 \includegraphics[width=7.5cm,clip]{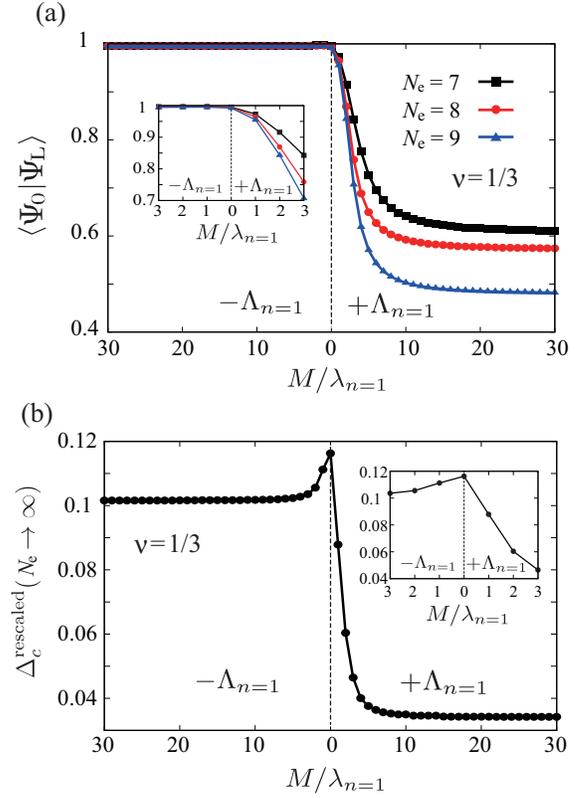}
 \caption{(Color online) (a) Mass dependence of the overlaps between the numerical ground state and the Laughlin wave function in $N_{e} = 7, 8, $ and $9$ systems. (b) Charge gap in thermodynamic limit with unit of $e^{2}/\epsilon l'_{B}$. 
Insets in (a) and (b) give the numerical results in $0 \le M/\lambda_{n=1} \le 3$}
 \label{fig:gap_overlap}

\end{figure}

We investigate the mass effect on the  $n=1$ positive and negative relativistic Landau levels. 
In particular, we discuss the mass dependent behavior of the Laughlin state of $n=1$, which is {the many-body groundstate} in the massless case. 
Figures \ref{fig:ppM}(a) and \ref{fig:ppM}(b) show the mass dependence of the pseudopotentials. 
{[To express the mass dependence of quantities,  we adopt a dimensionless mass parameter, $M/\lambda_{n=1}$ with $\lambda_{n=1}$ (\ref{eigenvaluesofdirac}) being the kinetic energy of the $n=1$ relativistic Landau level.]}  
As shown in Fig.~\ref{fig:ppM}(a), for the negative relativistic Landau level $-\Lambda_{n=1} < 0$,  $V^{\rm R}_{m}$ slightly decrease with the increase of $M/\lambda_{n=1}$ and  monotonically  with the increase of  the relative angular momentum $m$ 
{like the $n=0$ $\it{non}$-relativistic pseudopotential.}   
For the positive Landau level $+\Lambda_{n=1}$, the pseudopotential exhibits an intriguing behavior: for small $M/\lambda_{n=1}$, its behavior is similar to that of the massless case in Fig.~\ref{fig:pp}(a), while with the increase of $M/\lambda_{n=1}$, the pseudopotential shows a cusp at $m=2$, which is a  unique feature of the $n=1$ $\it{non}$-relativistic pseudopotential as depicted in Fig.~\ref{fig:pp}(b). 
{We thus have derived a concrete behavior of the pseudopotentials with respect to the mass parameter and  demonstrated that the mass parameter  interpolates the relativistic pseudo-potential $(M\rightarrow 0)$ and  non-relativistic one $(M\rightarrow \infty)$. }  

{Since the pseudo-potential Hamiltonian dominates the many-body physics in a given Landau level, it may be natural to expect that  the mass  parameter also interpolates the relativistic  and non-relativistic many-body physics. }  
The overlap $\langle \Psi_{0} | \Psi_{\rm L} \rangle$ {at $\nu=1/3$} in $N_{e}=7,8$ and $9$ systems as a function of $M/\lambda_{n=1}$ is shown in Fig.~\ref{fig:gap_overlap}(a). 
The overlap in  the negative relativistic Landau level $-\Lambda_{n=1}$ keeps  large values more than $99 \%$ all over $M/\lambda_{n=1}$ (left-figure). 
Meanwhile, the overlap in the relativistic positive Landau level $+\Lambda_{n=1}$ rapidly decreases with increasing $M/\lambda_{n=1}$ (right-figure) and finally reduces to about $50\%$ in a $N_{e} = 9$ system in the large $M/\lambda_{n=1}$ region.  
{Besides, the overlap becomes strongly suppressed with  the increase of $N_e$ [Fig.~\ref{fig:gap_overlap}(a)]. Hence in the thermodynamic limit, the overlap will be  significantly reduced even if  the mass is not so large.}    
{The charge gap in the thermodynamic limit $(N_e\rightarrow \infty)$  can be obtained from the  rescaled charge gap $\Delta_c^{\text{rescaled}}$  as deduced at  $1/N_e\rightarrow 0$ in Fig.\ref{fig:energy}(d).} 
The  charge gap $\Delta_c^{\text{rescaled}}(N_e\rightarrow \infty)$ also exhibits a rapid decay in $+\Lambda_{n=1}$ as shown in Fig.~\ref{fig:gap_overlap}(b). 
Thus the Laughlin state is no longer a good candidate for the ground state with large $M/\lambda_{n=1}$ in $+\Lambda_{n}$, {and in the $M\rightarrow\infty$ limit the groundstate will become to that of the non-relativistic $n=1$  Landau level  \cite{Shibata-Yoshioka-2003}. }

Let us consider these results in view of the parity anomaly. 
In the massless case, the relativistic system respects the chiral symmetry and so the pseudopotentials in the  positive and negative relativistic Landau levels are equivalent.   
When $M$ is turned on,  accompanied with the chiral symmetry breaking,  the mass deformed pseudopotentials in $\pm\Lambda_{n}$ begin to split according to (\ref{eq:ppM}). 
Since in the non-relativistic limit $(M \rightarrow \infty)$,   $\psi_{+\Lambda_{n}}$  and $\psi_{-\Lambda_{n}}$  are reduced to  the non-relativistic eigenstates in $n$th and $(n-1)$th Landau levels [remember the discussions in Sect.\ref{subsec:massdeformation}],   
 $V^{\rm R}_{J}(M)_{+\Lambda_{n}}$ and $V^{\rm R}_{J}(M)_{-\Lambda_{n}}$ are  also reduced to  the non-relativistic pseudopotentials of the  $n$ and $(n-1)$th Landau levels respectively.  
As a result, while the ground state in $-\Lambda_{n=1}$ keeps being characterized by the Laughlin state,  { in $+\Lambda_{n=1}$  the groundstate is no longer given by the  Laughlin state in the large $M$ region.} 
Thus, the ``parity anomaly'' brings inequivalence also to the many-body physics of the relativistic quantum Hall effect.  

\section{Summary and discussions}

We developed a relativistic formulation of the quantum Hall effect on the Haldane sphere and  performed  numerical investigations. 
Specifically, we analyzed the properties of the relativistic quantum Hall liquid at $\nu = {1}/{3}$ by the exact diagonalization
 based on the newly constructed  relativistic pseudopotential Hamiltonian.      
 Though in the massless case either of relativistic many-body groundstates for the $n=1$ positive and negative relativistic Landau levels are well described by the Laughlin wavefunction,  our numerical results indicate that {the mass term significantly reduces the charge gap} of the positive relativistic  Landau level $n = 1$ and the Laughlin state  no longer well describes the many-body groundstate in the presence of a sufficiently large mass.  
{Since the mass-dependent behaviors are enhanced with the increase of the number of electrons,   our results will maintain correct in the thermodynamic limit even if the mass gap is not so large.}   

Here, we note the energy scale of the mass term in experimental Dirac matter. { For graphene  and topological insulators (such as ${\rm Bi}_{2}{\rm Se}_{3}$), the Fermi velocity $v_F$ is  in the order of $1 \times 10^{6}$ \cite{Sarma-et-al-2011} and $1 \times 10^{5}$ m/s \cite{Cheng-et-al-2010}, respectively.   The  level spacing between the relativistic $n=0$ and $n=1$ Landau levels is given by $\sqrt{2}\hbar v_F/\ell_B\simeq 4v_{\rm F}\sqrt{B[{\rm T}]} \times 10^{-5}$ meV.   
Meanwhile,  the mass gap induced by the breaking of the AB sublattice symmetry of graphene is reported as  $260$ meV \cite{Zhou-et-al-2007}, and 
that of the ${\rm Bi}_{2}{\rm Se}_{3}$ topological insulator surface  with $16 \%$ Fe doping is estimated about $50$ meV \cite{Chen-et-al-2010}.} 
Thus, for both graphene and topological insulator surfaces, the mass parameter can become  comparable with (or even larger than)
 the Landau level spacing  by controlling an external magnetic field. 
It is expected that the present results will be observable  in  real Dirac matter.

\section*{Note~Added in Proof}

{
 While  revising the paper, we were aware of a work by Arciniaga and Peterson \cite{Arciniaga-Peterson-2016} in which they independently
  explored a relativistic formulation of the quantum Hall effect on the Haldane
  sphere. 
} 
 
\begin{acknowledgments}

KH would like to thank Hiroaki Matsueda for arranging a seminar at  Tohoku University  which enabled the present collaboration. 
This work was supported by {
JSPS KAKENHI Grant No. 26400344, No. 16K05334, and No. 16K05138.}
\end{acknowledgments}
\appendix
  \makeatletter

    \renewcommand{\theequation}{%
    \Alph{section}.\arabic{equation}}
    \@addtoreset{equation}{section}
  \makeatother


\section{ \label{seq:drv_pp} Derivation of the pseudopotential}

We derive an explicit representation of $V_{J}^{\rm R}$ (\ref{relativepseudopo}) using the method of Ref.\cite{Wooten-Macek-2014}.  
From the definition of the pseudopotential and the two-particle state,  $V(j,\alpha,\beta)$ in the main text is given by 
\begin{align}
V_J&(j,\alpha,\beta) = \nonumber \\
&\  \sum_{m_{1},m_{2},n_{1},n_{2}} C^{J,J_{z}}_{jm_{1},jm_{2}} C^{J,J_{z}}_{jn_{1},jn_{2}} \nonumber \\
&\ \langle j,m_{1},\alpha;j,m_{2},\beta | V | j,n_{1},\alpha;j,n_{2},\beta \rangle.   
\label{exppseudopotapp}
\end{align}
Here, $V$ is the Coulomb interaction on the sphere,  {$V(\boldsymbol{r}_{1}-\boldsymbol{r}_{2})=\frac{e^2}{\epsilon|\boldsymbol{r}_{1}-\boldsymbol{r}_{2}|}$ with $|\boldsymbol{r}_1|=|\boldsymbol{r}_2|=R$,}  which can be expressed as 
\begin{equation}
V(\boldsymbol{r}_{1}-\boldsymbol{r}_{2}) = \frac{ e^{2} }{\epsilon R } \sum_{k=0}^{\infty} \frac{ 4\pi }{2k+1} \sum_{m'=-k}^{k}  Y_{k,m'}(\Omega_{1})^*  Y_{k,m'}(\Omega_{2})
\end{equation}
where $Y_{k,m'}$ are spherical harmonics and $\Omega_{p}$ means the spherical coordinate ($\theta_{p}, \phi_{p}$) for the $p$th particle. The matrix elements of (\ref{exppseudopotapp})  are written as 
\begin{align}
\langle j,&m_{1},\alpha;j,m_{2},\beta | V | j,n_{1},\alpha;j,n_{2},\beta \rangle = \nonumber \\
&\frac{ e^{2} }{ \epsilon R } \sum_{k=0}^{\infty} \sum_{m'=-k}^{k} \frac{ 4\pi }{2k+1} \nonumber \\
&\times \int  Y^{\alpha}_{j,m_{1}} (\Omega_{1})^{*}  Y_{k,m'} (\Omega_{1})^{*} Y^{\alpha}_{j,n_{1}} (\Omega_{1})  d \Omega_{1} \nonumber \\
&\times \int  Y^{\beta}_{j,m_{2}} (\Omega_{2})^{*}  Y_{k,m'} (\Omega_{2}) Y^{\beta}_{j,n_{2}} (\Omega_{1}) d \Omega_{2}. \label{mateleofv}
\end{align}
Using properties of the monopole harmonics 
 \cite{Wu-Yang-1976,Wu-Yang-1977} 
\begin{subequations}
\begin{align}
&Y_{k,m'}(\Omega) = Y^{0}_{k,m'}(\Omega) \\ 
&Y^{g}_{j,m}(\Omega)^{*} = (-1)^{g+M} Y^{-g}_{j,-m}(\Omega) \\
&\int Y^{g}_{j,m}(\Omega) Y^{0}_{k,m'}(\Omega) Y^{g'}_{j,n}(\Omega) d \Omega = (-1)^{2j+k} \times \nonumber \\
& \left[ \frac{ (2j+1)(2k+1)(2j'+1) }{ 4 \pi } \right]^{1/2} \left( 
 \begin{array}{ccc}
   j & k & j \\
  g & 0 & g'
 \end{array}
 \right) \left( 
 \begin{array}{ccc}
   j & k & j \\
  m & m' & n
 \end{array}
 \right), 
\end{align}
\end{subequations}
we can represent (\ref{mateleofv}) as 
\begin{align}
\langle j,&m_{1},\alpha;j,m_{2},\beta | V | j,n_{1},\alpha;j,n_{2},\beta \rangle = \nonumber \\
&\frac{ e^{2} }{ \epsilon R } \sum_{k=0}^{2j} \sum_{m'=-k}^{k} (2j+1)^{2}
\left( 
 \begin{array}{ccc}
   j & k & j \\
  -\alpha & 0 & \alpha
 \end{array}
 \right) \left( 
 \begin{array}{ccc}
   j & k & j \\
  -\beta & 0 & \beta
 \end{array}
 \right) \nonumber \\
&\times \left( 
 \begin{array}{ccc}
   j & k & j \\
  -m_{1} & -m' & n_{1}
 \end{array}
 \right) \left( 
 \begin{array}{ccc}
   j & k & j \\
  -m_{2} & m' & n_{2}
 \end{array}
 \right) . 
\end{align} 
Note that $\sum_{k=0}^{\infty}$ reduces a finite sum up to $2j$ since the 3j-symbol is finite for $0 \le k \le 2j$.	
{Since $V_J(j, \alpha, \beta)$ does not depend on $J_z$, the sum on the right-hand side of (\ref{exppseudopotapp}) can be rewritten as    
\be
\sum_{m_1, m_2, n_1, n_2, m'}(\cdots)=\frac{1}{2J+1}\sum_{m_1, m_2, n_1, n_2, m', J_z}(\cdots).   
\ee
With (\ref{defCG3j}) and the definition of the 6j-symbol  
\begin{align}
\begin{Bmatrix}
j_1 & j_2 & j_3 \\
j_4 & j_5 & j_6 
\end{Bmatrix}
 & \equiv \sum_{\{m_i\}} 
(-1)^{\sum_{i=1}^6 (j_i-m_i)} 
\nn\\ 
&\times \begin{pmatrix}
j_1 & j_2 & j_3 \\
-m_1 & -m_2 & -m_3
\end{pmatrix} \begin{pmatrix}
j_1 & j_5 & j_6 \\
m_1 & -m_5 & m_6
\end{pmatrix}  \nn\\
&\times \begin{pmatrix}
j_4 & j_2 & j_6 \\
m_4 & m_2 & -m_6
\end{pmatrix} \begin{pmatrix}
j_4 & j_5 & j_3 \\
-m_4 & m_5 & m_3
\end{pmatrix}, 
\end{align}
we  can perform the summation  over $m_1, m_2, n_1, n_2, m', J_z$  in (\ref{exppseudopotapp}) to  obtain (\ref{comppseudopoterela}).  } 

\appendix

\nocite{*}

\providecommand{\noopsort}[1]{}\providecommand{\singleletter}[1]{#1}%


\begin{thebibliography}{33}%
\makeatletter
\providecommand \@ifxundefined [1]{%
 \@ifx{#1\undefined}
}%
\providecommand \@ifnum [1]{%
 \ifnum #1\expandafter \@firstoftwo
 \else \expandafter \@secondoftwo
 \fi
}%
\providecommand \@ifx [1]{%
 \ifx #1\expandafter \@firstoftwo
 \else \expandafter \@secondoftwo
 \fi
}%
\providecommand \natexlab [1]{#1}%
\providecommand \enquote  [1]{``#1''}%
\providecommand \bibnamefont  [1]{#1}%
\providecommand \bibfnamefont [1]{#1}%
\providecommand \citenamefont [1]{#1}%
\providecommand \href@noop [0]{\@secondoftwo}%
\providecommand \href [0]{\begingroup \@sanitize@url \@href}%
\providecommand \@href[1]{\@@startlink{#1}\@@href}%
\providecommand \@@href[1]{\endgroup#1\@@endlink}%
\providecommand \@sanitize@url [0]{\catcode `\\12\catcode `\$12\catcode
  `\&12\catcode `\#12\catcode `\^12\catcode `\_12\catcode `\%12\relax}%
\providecommand \@@startlink[1]{}%
\providecommand \@@endlink[0]{}%
\providecommand \url  [0]{\begingroup\@sanitize@url \@url }%
\providecommand \@url [1]{\endgroup\@href {#1}{\urlprefix }}%
\providecommand \urlprefix  [0]{URL }%
\providecommand \Eprint [0]{\href }%
\providecommand \doibase [0]{http://dx.doi.org/}%
\providecommand \selectlanguage [0]{\@gobble}%
\providecommand \bibinfo  [0]{\@secondoftwo}%
\providecommand \bibfield  [0]{\@secondoftwo}%
\providecommand \translation [1]{[#1]}%
\providecommand \BibitemOpen [0]{}%
\providecommand \bibitemStop [0]{}%
\providecommand \bibitemNoStop [0]{.\EOS\space}%
\providecommand \EOS [0]{\spacefactor3000\relax}%
\providecommand \BibitemShut  [1]{\csname bibitem#1\endcsname}%
\let\auto@bib@innerbib\@empty
\bibitem [{\citenamefont {Wallece}(1947)}]{Wallace-1947}%
  \BibitemOpen
  \bibfield  {author} {\bibinfo {author} {\bibfnamefont {P.~R.}\ \bibnamefont
  {Wallece}},\ }\href@noop {} {\bibfield  {journal} {\bibinfo  {journal} {Phys.
  Rev.}\ }\textbf {\bibinfo {volume} {71}},\ \bibinfo {pages} {622} (\bibinfo
  {year} {1947})}\BibitemShut {NoStop}%
\bibitem [{\citenamefont {Zhou}\ \emph {et~al.}(2006)\citenamefont {Zhou},
  \citenamefont {Gweon}, \citenamefont {Graf}, \citenamefont {Fedorov},
  \citenamefont {Spataru}, \citenamefont {Diehl}, \citenamefont {Kopelevich},
  \citenamefont {Lee}, \citenamefont {Louie},\ and\ \citenamefont
  {Lanzara}}]{Zhou-et-al-2006}%
  \BibitemOpen
  \bibfield  {author} {\bibinfo {author} {\bibfnamefont {S.~Y.}\ \bibnamefont
  {Zhou}}, \bibinfo {author} {\bibfnamefont {G.-H.}\ \bibnamefont {Gweon}},
  \bibinfo {author} {\bibfnamefont {J.}~\bibnamefont {Graf}}, \bibinfo {author}
  {\bibfnamefont {A.~V.}\ \bibnamefont {Fedorov}}, \bibinfo {author}
  {\bibfnamefont {C.~D.}\ \bibnamefont {Spataru}}, \bibinfo {author}
  {\bibfnamefont {R.~D.}\ \bibnamefont {Diehl}}, \bibinfo {author}
  {\bibfnamefont {Y.}~\bibnamefont {Kopelevich}}, \bibinfo {author}
  {\bibfnamefont {D.-H.}\ \bibnamefont {Lee}}, \bibinfo {author} {\bibfnamefont
  {S.~G.}\ \bibnamefont {Louie}}, \ and\ \bibinfo {author} {\bibfnamefont
  {A.}~\bibnamefont {Lanzara}},\ }\href@noop {} {\bibfield  {journal} {\bibinfo
   {journal} {Nature physics}\ }\textbf {\bibinfo {volume} {2}},\ \bibinfo
  {pages} {595} (\bibinfo {year} {2006})}\BibitemShut {NoStop}%
\bibitem [{\citenamefont {Bostwick}\ \emph {et~al.}(2007)\citenamefont
  {Bostwick}, \citenamefont {Ohta}, \citenamefont {Seyller}, \citenamefont
  {Horn},\ and\ \citenamefont {Rotenberg}}]{Bostwick-et-al-2007}%
  \BibitemOpen
  \bibfield  {author} {\bibinfo {author} {\bibfnamefont {A.}~\bibnamefont
  {Bostwick}}, \bibinfo {author} {\bibfnamefont {T.}~\bibnamefont {Ohta}},
  \bibinfo {author} {\bibfnamefont {T.}~\bibnamefont {Seyller}}, \bibinfo
  {author} {\bibfnamefont {K.}~\bibnamefont {Horn}}, \ and\ \bibinfo {author}
  {\bibfnamefont {E.}~\bibnamefont {Rotenberg}},\ }\href@noop {} {\bibfield
  {journal} {\bibinfo  {journal} {Nature physics}\ }\textbf {\bibinfo {volume}
  {3}},\ \bibinfo {pages} {36} (\bibinfo {year} {2007})}\BibitemShut {NoStop}%
\bibitem [{\citenamefont {Zhang}\ \emph {et~al.}(2009)\citenamefont {Zhang},
  \citenamefont {Liu}, \citenamefont {Qi}, \citenamefont {Dai}, \citenamefont
  {Fang},\ and\ \citenamefont {Zhang}}]{Zhang-et-al-2009}%
  \BibitemOpen
  \bibfield  {author} {\bibinfo {author} {\bibfnamefont {H.}~\bibnamefont
  {Zhang}}, \bibinfo {author} {\bibfnamefont {C.-X.}\ \bibnamefont {Liu}},
  \bibinfo {author} {\bibfnamefont {X.-L.}\ \bibnamefont {Qi}}, \bibinfo
  {author} {\bibfnamefont {Xi}~\bibnamefont {Dai}}, \bibinfo {author}
  {\bibfnamefont {Z.}~\bibnamefont {Fang}}, \ and\ \bibinfo {author}
  {\bibfnamefont {S.-C.}\ \bibnamefont {Zhang}},\ }\href@noop {} {\bibfield
  {journal} {\bibinfo  {journal} {Nature physics}\ }\textbf {\bibinfo {volume}
  {5}},\ \bibinfo {pages} {438} (\bibinfo {year} {2009})}\BibitemShut {NoStop}%
\bibitem [{\citenamefont {Zhang}\ \emph {et~al.}(2005)\citenamefont {Zhang},
  \citenamefont {Tan}, \citenamefont {Stormer},\ and\ \citenamefont
  {Kim}}]{Zhang-et-al-2005}%
  \BibitemOpen
  \bibfield  {author} {\bibinfo {author} {\bibfnamefont {Y.}~\bibnamefont
  {Zhang}}, \bibinfo {author} {\bibfnamefont {Y.-W.}\ \bibnamefont {Tan}},
  \bibinfo {author} {\bibfnamefont {H.~L.}\ \bibnamefont {Stormer}}, \ and\
  \bibinfo {author} {\bibfnamefont {P.}~\bibnamefont {Kim}},\ }\href@noop {}
  {\bibfield  {journal} {\bibinfo  {journal} {Nature}\ }\textbf {\bibinfo
  {volume} {438}},\ \bibinfo {pages} {201} (\bibinfo {year}
  {2005})}\BibitemShut {NoStop}%
\bibitem [{\citenamefont {Novoselov}\ \emph {et~al.}(2005)\citenamefont
  {Novoselov}, \citenamefont {Geim}, \citenamefont {Morozov}, \citenamefont
  {Jiang}, \citenamefont {Katsnelson}, \citenamefont {Grigorieva},
  \citenamefont {Dubonos},\ and\ \citenamefont
  {Firsov}}]{Novoselov-et-al-2005}%
  \BibitemOpen
  \bibfield  {author} {\bibinfo {author} {\bibfnamefont {K.~S.}\ \bibnamefont
  {Novoselov}}, \bibinfo {author} {\bibfnamefont {A.~K.}\ \bibnamefont {Geim}},
  \bibinfo {author} {\bibfnamefont {S.~V.}\ \bibnamefont {Morozov}}, \bibinfo
  {author} {\bibfnamefont {D.}~\bibnamefont {Jiang}}, \bibinfo {author}
  {\bibfnamefont {M.~I.}\ \bibnamefont {Katsnelson}}, \bibinfo {author}
  {\bibfnamefont {I.~V.}\ \bibnamefont {Grigorieva}}, \bibinfo {author}
  {\bibfnamefont {S.~V.}\ \bibnamefont {Dubonos}}, \ and\ \bibinfo {author}
  {\bibfnamefont {A.~A.}\ \bibnamefont {Firsov}},\ }\href@noop {} {\bibfield
  {journal} {\bibinfo  {journal} {Nature}\ }\textbf {\bibinfo {volume} {438}},\
  \bibinfo {pages} {197} (\bibinfo {year} {2005})}\BibitemShut {NoStop}%
\bibitem [{\citenamefont {Bolotin}\ \emph {et~al.}(2009)\citenamefont
  {Bolotin}, \citenamefont {Ghahari}, \citenamefont {Shulman}, \citenamefont
  {Stormer},\ and\ \citenamefont {Kim}}]{Bolotin-et-al-2011}%
  \BibitemOpen
  \bibfield  {author} {\bibinfo {author} {\bibfnamefont {K.~I.}\ \bibnamefont
  {Bolotin}}, \bibinfo {author} {\bibfnamefont {F.}~\bibnamefont {Ghahari}},
  \bibinfo {author} {\bibfnamefont {M.~D.}\ \bibnamefont {Shulman}}, \bibinfo
  {author} {\bibfnamefont {H.~L.}\ \bibnamefont {Stormer}}, \ and\ \bibinfo
  {author} {\bibfnamefont {P.}~\bibnamefont {Kim}},\ }\href@noop {} {\bibfield
  {journal} {\bibinfo  {journal} {Nature Phys.}\ }\textbf {\bibinfo {volume}
  {462}},\ \bibinfo {pages} {196} (\bibinfo {year} {2009})}\BibitemShut
  {NoStop}%
\bibitem [{\citenamefont {Yoshimi}\ \emph {et~al.}(2015)\citenamefont
  {Yoshimi}, \citenamefont {Tsukazaki}, \citenamefont {Kozuka}, \citenamefont
  {Falson}, \citenamefont {Takahashi}, \citenamefont {Checkelsky},
  \citenamefont {Nagaosa}, \citenamefont {Kawasaki},\ and\ \citenamefont
  {Tokura}}]{Yoshimi-et-al-2015}%
  \BibitemOpen
  \bibfield  {author} {\bibinfo {author} {\bibfnamefont {R.}~\bibnamefont
  {Yoshimi}}, \bibinfo {author} {\bibfnamefont {A.}~\bibnamefont {Tsukazaki}},
  \bibinfo {author} {\bibfnamefont {Y.}~\bibnamefont {Kozuka}}, \bibinfo
  {author} {\bibfnamefont {J.}~\bibnamefont {Falson}}, \bibinfo {author}
  {\bibfnamefont {K.~S.}~\bibnamefont {Takahashi}}, \bibinfo {author}
  {\bibfnamefont {J.~G.}~\bibnamefont {Checkelsky}}, \bibinfo {author}
  {\bibfnamefont {N.}~\bibnamefont {Nagaosa}}, \bibinfo {author} {\bibfnamefont
  {M.}~\bibnamefont {Kawasaki}}, \ and\ \bibinfo {author} {\bibfnamefont
  {Y.}~\bibnamefont {Tokura}},\ }\href@noop {} {\bibfield  {journal} {\bibinfo
  {journal} {Nature communications}\ }\textbf {\bibinfo {volume} {6}},\
  \bibinfo {pages} {6627} (\bibinfo {year} {2015})}\BibitemShut {NoStop}%
\bibitem [{\citenamefont {Xu}\ \emph {et~al.}(2014)\citenamefont {Xu},
  \citenamefont {Miotkowski}, \citenamefont {Liu}, \citenamefont {Tian},
  \citenamefont {Nam}, \citenamefont {Alidoust}, \citenamefont {Hu},
  \citenamefont {Shih}, \citenamefont {Hasan},\ and\ \citenamefont
  {Chen}}]{Xue-et-al-2014}%
  \BibitemOpen
  \bibfield  {author} {\bibinfo {author} {\bibfnamefont {Y.}~\bibnamefont
  {Xu}}, \bibinfo {author} {\bibfnamefont {I.}~\bibnamefont {Miotkowski}},
  \bibinfo {author} {\bibfnamefont {C.}~\bibnamefont {Liu}}, \bibinfo {author}
  {\bibfnamefont {J.}~\bibnamefont {Tian}}, \bibinfo {author} {\bibfnamefont
  {H.}~\bibnamefont {Nam}}, \bibinfo {author} {\bibfnamefont {N.}~\bibnamefont
  {Alidoust}}, \bibinfo {author} {\bibfnamefont {J.}~\bibnamefont {Hu}},
  \bibinfo {author} {\bibfnamefont {C.~K.}\ \bibnamefont {Shih}}, \bibinfo
  {author} {\bibfnamefont {M.~Z.}\ \bibnamefont {Hasan}}, \ and\ \bibinfo
  {author} {\bibfnamefont {Y.~P.}\ \bibnamefont {Chen}},\ }\href@noop {}
  {\bibfield  {journal} {\bibinfo  {journal} {Nature Physics}\ }\textbf
  {\bibinfo {volume} {10}},\ \bibinfo {pages} {956} (\bibinfo {year}
  {2014})}\BibitemShut {NoStop}%
\bibitem [{\citenamefont {Qi}\ \emph {et~al.}(2008)\citenamefont {Qi},
  \citenamefont {Hughes},\ and\ \citenamefont {Zhang}}]{QiHZ2008}%
  \BibitemOpen
  \bibfield  {author} {\bibinfo {author} {\bibfnamefont {X.-L.}\ \bibnamefont
  {Qi}}, \bibinfo {author} {\bibfnamefont {T.~L.}~\bibnamefont {Hughes}}, \ and\
  \bibinfo {author} {\bibfnamefont {S.-C.}\ \bibnamefont {Zhang}},\ }\href@noop
  {} {\bibfield  {journal} {\bibinfo  {journal} {Phys. Rev. B}\ }\textbf
  {\bibinfo {volume} {78}},\ \bibinfo {pages} {195424} (\bibinfo {year}
  {2008})}\BibitemShut {NoStop}%
\bibitem [{\citenamefont {Zhou}\ \emph {et~al.}(2007)\citenamefont {Zhou},
  \citenamefont {Gweon}, \citenamefont {Fedorov}, \citenamefont {First},
  \citenamefont {de~Heer}, \citenamefont {Lee}, \citenamefont {Guinea},
  \citenamefont {Neto},\ and\ \citenamefont {Lanzara}}]{Zhou-et-al-2007}%
  \BibitemOpen
  \bibfield  {author} {\bibinfo {author} {\bibfnamefont {S.~Y.}\ \bibnamefont
  {Zhou}}, \bibinfo {author} {\bibfnamefont {G.-H.}\ \bibnamefont {Gweon}},
  \bibinfo {author} {\bibfnamefont {A.~V.}\ \bibnamefont {Fedorov}}, \bibinfo
  {author} {\bibfnamefont {P.~N.}\ \bibnamefont {First}}, \bibinfo {author}
  {\bibfnamefont {W.~A.}\ \bibnamefont {de~Heer}}, \bibinfo {author}
  {\bibfnamefont {D.-H.}\ \bibnamefont {Lee}}, \bibinfo {author} {\bibfnamefont
  {F.}~\bibnamefont {Guinea}}, \bibinfo {author} {\bibfnamefont {A.~H.}\
  \bibnamefont {Castro~Neto}}, \ and\ \bibinfo {author} {\bibfnamefont
  {A.}~\bibnamefont {Lanzara}},\ }\href@noop {} {\bibfield  {journal} {\bibinfo
   {journal} {Nature Materials}\ }\textbf {\bibinfo {volume} {6}},\ \bibinfo
  {pages} {770} (\bibinfo {year} {2007})}\BibitemShut {NoStop}%
\bibitem [{\citenamefont {Pereira}\ \emph {et~al.}(2008)\citenamefont
  {Pereira}, \citenamefont {dos Santos},\ and\ \citenamefont
  {Neto}}]{Pereira-et-al-2008}%
  \BibitemOpen
  \bibfield  {author} {\bibinfo {author} {\bibfnamefont {V.~M.}\ \bibnamefont
  {Pereira}}, \bibinfo {author} {\bibfnamefont {J.~M.~B.}\ \bibnamefont {Lopes~dos~Santos}}, \ and\ \bibinfo {author} {\bibfnamefont {A.~H.}\ \bibnamefont
  {Castro~Neto}},\ }\href@noop {} {\bibfield  {journal} {\bibinfo  {journal}
  {Phys.~Rev.~B}\ }\textbf {\bibinfo {volume} {77}},\ \bibinfo {pages} {115109}
  (\bibinfo {year} {2008})}\BibitemShut {NoStop}%
\bibitem [{\citenamefont {Chen}\ \emph {et~al.}(2010)\citenamefont {Chen},
  \citenamefont {Chu}, \citenamefont {Analytis}, \citenamefont {Liu},
  \citenamefont {Igarashi}, \citenamefont {Kuo}, \citenamefont {Qi},
  \citenamefont {Mo}, \citenamefont {Moore}, \citenamefont {Lu}, \citenamefont
  {Hashimoto}, \citenamefont {Sasagawa}, \citenamefont {Zhang}, \citenamefont
  {Fisher}, \citenamefont {Jussain},\ and\ \citenamefont
  {Shin}}]{Chen-et-al-2010}%
  \BibitemOpen
  \bibfield  {author} {\bibinfo {author} {\bibfnamefont {Y.~L.}\ \bibnamefont
  {Chen}}, \bibinfo {author} {\bibfnamefont {J.-H.}\ \bibnamefont {Chu}},
  \bibinfo {author} {\bibfnamefont {J.~G.}\ \bibnamefont {Analytis}}, \bibinfo
  {author} {\bibfnamefont {X.~K.}\ \bibnamefont {Liu}}, \bibinfo {author}
  {\bibfnamefont {K.}~\bibnamefont {Igarashi}}, \bibinfo {author}
  {\bibfnamefont {H.-H.}\ \bibnamefont {Kuo}}, \bibinfo {author} {\bibfnamefont
  {X.~L.}\ \bibnamefont {Qi}}, \bibinfo {author} {\bibfnamefont {S.~K.}\
  \bibnamefont {Mo}}, \bibinfo {author} {\bibfnamefont {R.~G.}\ \bibnamefont
  {Moore}}, \bibinfo {author} {\bibfnamefont {D.~H.}\ \bibnamefont {Lu}},
  \bibinfo {author} {\bibfnamefont {M.}~\bibnamefont {Hashimoto}}, \bibinfo
  {author} {\bibfnamefont {T.}~\bibnamefont {Sasagawa}}, \bibinfo {author}
  {\bibfnamefont {S.~C.}\ \bibnamefont {Zhang}}, \bibinfo {author}
  {\bibfnamefont {I.~R.}\ \bibnamefont {Fisher}}, \bibinfo {author}
  {\bibfnamefont {X.}~\bibnamefont {Hussain}}, \ and\ \bibinfo {author}
  {\bibfnamefont {Z.~X.}\ \bibnamefont {Shen}},\ }\href@noop {} {\bibfield
  {journal} {\bibinfo  {journal} {Science}\ }\textbf {\bibinfo {volume}
  {329}},\ \bibinfo {pages} {659} (\bibinfo {year} {2010})}\BibitemShut
  {NoStop}%
\bibitem [{Note1()}]{Note1}%
  \BibitemOpen
  \bibinfo {note} {The parity anomaly is usually referred to as the parity
  breaking Chern-Simons term induced by quantum effect. In the present paper
  the ``parity anomaly'' is referred to as asymmetry of the positive and
  negative Landau levels due to the (parity breaking) mass term in
  (2+1)D.}\BibitemShut {Stop}%
\bibitem [{Note2()}]{Note2}%
  \BibitemOpen
  \bibinfo {note} {The mass term does not necessarily violate the reflection
  symmetry of the Dirac operator spectrum with respect to the zero energy. For
  instance in the case of the free Dirac operator, the mass deformation breaks
  the chiral symmetry, but the spectrum still preserves the reflection
  symmetry. The chiral symmetry is a sufficient condition of the reflection
  symmetry of the spectrum but not a necessary condition.}\BibitemShut {Stop}%
\bibitem [{\citenamefont {Haldane}(1988)}]{Haldane1988}%
  \BibitemOpen
  \bibfield  {author} {\bibinfo {author} {\bibfnamefont {F.~D.~M.}\
  \bibnamefont {Haldane}},\ }\href@noop {} {\bibfield  {journal} {\bibinfo
  {journal} {Phys.\ Rev.\ Lett.}\ }\textbf {\bibinfo {volume} {61}},\ \bibinfo
  {pages} {2015} (\bibinfo {year} {1988})}\BibitemShut {NoStop}%
\bibitem [{\citenamefont {Haldane}(1983)}]{Haldane1983}%
  \BibitemOpen
  \bibfield  {author} {\bibinfo {author} {\bibfnamefont {F.~D.~M.}\
  \bibnamefont {Haldane}},\ }\href@noop {} {\bibfield  {journal} {\bibinfo
  {journal} {Phys.\ Rev.\ Lett.}\ }\textbf {\bibinfo {volume} {51}},\ \bibinfo
  {pages} {605} (\bibinfo {year} {1983})}\BibitemShut {NoStop}%
\bibitem [{\citenamefont {Shibata}\ and\ \citenamefont
  {Nomura}(2009)}]{Shibata-Nomura-2009}%
  \BibitemOpen
  \bibfield  {author} {\bibinfo {author} {\bibfnamefont {N.}~\bibnamefont
  {Shibata}}\ and\ \bibinfo {author} {\bibfnamefont {K.}~\bibnamefont
  {Nomura}},\ }\href@noop {} {\bibfield  {journal} {\bibinfo  {journal} {J.
  Phys. Soc. Jpn.}\ }\textbf {\bibinfo {volume} {78}},\ \bibinfo {pages}
  {104708} (\bibinfo {year} {2009})}\BibitemShut {NoStop}%
\bibitem [{\citenamefont {Hasebe}()}]{Hasebe-2015}%
  \BibitemOpen
  \bibfield  {author} {\bibinfo {author} {\bibfnamefont {K.}~\bibnamefont
  {Hasebe}},\ }\href@noop {} {\ }\Eprint {http://arxiv.org/abs/1511.04681}
  {arXiv:1511.04681} \BibitemShut {NoStop}%
\bibitem [{\citenamefont {T\H{o}ke}\ \emph {et~al.}(2006)\citenamefont
  {T\H{o}ke}, \citenamefont {Lammert}, \citenamefont {Crespi},\ and\
  \citenamefont {Jain}}]{Toke-et-al-2006}%
  \BibitemOpen
  \bibfield  {author} {\bibinfo {author} {\bibfnamefont {C.}~\bibnamefont
  {T\H{o}ke}}, \bibinfo {author} {\bibfnamefont {P.~E.}\ \bibnamefont
  {Lammert}}, \bibinfo {author} {\bibfnamefont {V.~H.}\ \bibnamefont {Crespi}},
  \ and\ \bibinfo {author} {\bibfnamefont {J.~K.}\ \bibnamefont {Jain}},\
  }\href@noop {} {\bibfield  {journal} {\bibinfo  {journal} {Phys.~Rev.~B}\
  }\textbf {\bibinfo {volume} {74}},\ \bibinfo {pages} {235417} (\bibinfo
  {year} {2006})}\BibitemShut {NoStop}%
\bibitem [{\citenamefont {Apalkov}\ and\ \citenamefont
  {Chacraborty}(2006)}]{Apalkov-2006}%
  \BibitemOpen
  \bibfield  {author} {\bibinfo {author} {\bibfnamefont {V.~M.}\ \bibnamefont
  {Apalkov}}\ and\ \bibinfo {author} {\bibfnamefont {T.}~\bibnamefont
  {Chakraborty}},\ }\href@noop {} {\bibfield  {journal} {\bibinfo  {journal}
  {Phys.~Rev.~Lett.}\ }\textbf {\bibinfo {volume} {97}},\ \bibinfo {pages}
  {126801} (\bibinfo {year} {2006})}\BibitemShut {NoStop}%
\bibitem [{\citenamefont {Wu}\ and\ \citenamefont {Yang}(1976)}]{Wu-Yang-1976}%
  \BibitemOpen
  \bibfield  {author} {\bibinfo {author} {\bibfnamefont {T.~T.}\ \bibnamefont
  {Wu}}\ and\ \bibinfo {author} {\bibfnamefont {C.~N.}\ \bibnamefont {Yang}},\
  }\href@noop {} {\bibfield  {journal} {\bibinfo  {journal} {Nucl.~Phys.~B}\
  }\textbf {\bibinfo {volume} {107}},\ \bibinfo {pages} {365} (\bibinfo {year}
  {1976})}\BibitemShut {NoStop}%
\bibitem [{\citenamefont {Wu}\ and\ \citenamefont {Yang}(1977)}]{Wu-Yang-1977}%
  \BibitemOpen
  \bibfield  {author} {\bibinfo {author} {\bibfnamefont {T.~T.}\ \bibnamefont
  {Wu}}\ and\ \bibinfo {author} {\bibfnamefont {C.~N.}\ \bibnamefont {Yang}},\
  }\href@noop {} {\bibfield  {journal} {\bibinfo  {journal} {Phys.~Rev.~D}\
  }\textbf {\bibinfo {volume} {16}},\ \bibinfo {pages} {1018} (\bibinfo {year}
  {1977})}\BibitemShut {NoStop}%
\bibitem [{\citenamefont {Newman}\ and\ \citenamefont
  {Penrose}(1966)}]{Newman-Penrose-1966}%
  \BibitemOpen
  \bibfield  {author} {\bibinfo {author} {\bibfnamefont {E.~T.}\ \bibnamefont
  {Newman}}\ and\ \bibinfo {author} {\bibfnamefont {R.}~\bibnamefont
  {Penrose}},\ }\href@noop {} {\bibfield  {journal} {\bibinfo  {journal} {J.
  Math. Phys.}\ }\textbf {\bibinfo {volume} {7}},\ \bibinfo {pages} {863}
  (\bibinfo {year} {1966})}\BibitemShut {NoStop}%
\bibitem [{\citenamefont {Laughlin}(1983)}]{Laughlin-1983}%
  \BibitemOpen
  \bibfield  {author} {\bibinfo {author} {\bibfnamefont {R.~B.}\ \bibnamefont
  {Laughlin}},\ }\href@noop {} {\bibfield  {journal} {\bibinfo  {journal}
  {Phys.~Rev.~Lett.}\ }\textbf {\bibinfo {volume} {50}},\ \bibinfo {pages}
  {1395} (\bibinfo {year} {1983})}\BibitemShut {NoStop}%
\bibitem [{\citenamefont {Haldane}\ and\ \citenamefont
  {Rezayi}(1985)}]{Haldane-Rezayi-1985}%
  \BibitemOpen
  \bibfield  {author} {\bibinfo {author} {\bibfnamefont {F.~D.~M.}\
  \bibnamefont {Haldane}}\ and\ \bibinfo {author} {\bibfnamefont {E.~H.}\
  \bibnamefont {Rezayi}},\ }\href@noop {} {\bibfield  {journal} {\bibinfo
  {journal} {Phys.~Rev.~Lett.}\ }\textbf {\bibinfo {volume} {54}},\ \bibinfo
  {pages} {237} (\bibinfo {year} {1985})}\BibitemShut {NoStop}%
\bibitem [{\citenamefont {Fano}\ \emph {et~al.}(1986)\citenamefont {Fano},
  \citenamefont {Ortolani},\ and\ \citenamefont {Colombo}}]{Fano-et-al-1986}%
  \BibitemOpen
  \bibfield  {author} {\bibinfo {author} {\bibfnamefont {G.}~\bibnamefont
  {Fano}}, \bibinfo {author} {\bibfnamefont {F.}~\bibnamefont {Ortolani}}, \
  and\ \bibinfo {author} {\bibfnamefont {E.}~\bibnamefont {Colombo}},\
  }\href@noop {} {\bibfield  {journal} {\bibinfo  {journal} {Phys.~Rev.~B}\
  }\textbf {\bibinfo {volume} {34}},\ \bibinfo {pages} {2670} (\bibinfo {year}
  {1986})}\BibitemShut {NoStop}%
\bibitem [{\citenamefont {Prange}\ and\ \citenamefont
  {Girvin}(1987)}]{Prange-Girvin-1987}%
  \BibitemOpen
  \bibfield  {author} {\bibinfo {author} {\bibfnamefont {R.~E.}\ \bibnamefont
  {Prange}}\ and\ \bibinfo {author} {\bibfnamefont {S.~M.}\ \bibnamefont
  {Girvin}},\ }\href@noop {} {\emph {\bibinfo {title} {The Quantum Hall
  Effect}}}\ (\bibinfo  {publisher} {Springer-Verlag,~New~York},\ \bibinfo
  {year} {1987})\BibitemShut {NoStop}%
\bibitem [{\citenamefont {Nomura}\ and\ \citenamefont
  {MacDonald}(2006)}]{Nomura-MacDonald-2006}%
  \BibitemOpen
  \bibfield  {author} {\bibinfo {author} {\bibfnamefont {K.}~\bibnamefont
  {Nomura}}\ and\ \bibinfo {author} {\bibfnamefont {A.~H.}\ \bibnamefont
  {MacDonald}},\ }\href@noop {} {\bibfield  {journal} {\bibinfo  {journal}
  {Phys. Rev. Lett.}\ }\textbf {\bibinfo {volume} {96}},\ \bibinfo {pages}
  {256602} (\bibinfo {year} {2006})}\BibitemShut {NoStop}%
\bibitem [{\citenamefont {Morf}\ \emph {et~al.}(2002)\citenamefont {Morf},
  \citenamefont {d'Ambrumenil},\ and\ \citenamefont {Sarma}}]{Morf-et-al-2002}%
  \BibitemOpen
  \bibfield  {author} {\bibinfo {author} {\bibfnamefont {R.~H.}\ \bibnamefont
  {Morf}}, \bibinfo {author} {\bibfnamefont {N.}~\bibnamefont {d'Ambrumenil}},
  \ and\ \bibinfo {author} {\bibfnamefont {S.}\ \bibnamefont {Das~Sarma}},\
  }\href@noop {} {\bibfield  {journal} {\bibinfo  {journal} {Phys.~Rev.~B}\
  }\textbf {\bibinfo {volume} {66}},\ \bibinfo {pages} {075408} (\bibinfo
  {year} {2002})}\BibitemShut {NoStop}%
\bibitem [{\citenamefont {Shibata}\ and\ \citenamefont
  {Yoshioka}(2003)}]{Shibata-Yoshioka-2003}%
  \BibitemOpen
  \bibfield  {author} {\bibinfo {author} {\bibfnamefont {N.}~\bibnamefont
  {Shibata}}\ and\ \bibinfo {author} {\bibfnamefont {D.}\ \bibnamefont
  {Yoshioka}},\ }\href@noop {} {\bibfield  {journal} {\bibinfo  {journal}
  {J.~Phys.~Soc.~Jpn.}\ }\textbf {\bibinfo {volume} {72}},\ \bibinfo {pages}
  {664} (\bibinfo {year} {2003})}\BibitemShut {NoStop}%
\bibitem [{\citenamefont {Sarma}\ \emph {et~al.}(2011)\citenamefont {Sarma},
  \citenamefont {Adam}, \citenamefont {Hwang},\ and\ \citenamefont
  {Rossi}}]{Sarma-et-al-2011}%
  \BibitemOpen
  \bibfield  {author} {\bibinfo {author} {\bibfnamefont {S.~D.}\ \bibnamefont
  {Sarma}}, \bibinfo {author} {\bibfnamefont {S.}~\bibnamefont {Adam}},
  \bibinfo {author} {\bibfnamefont {E.~H.}\ \bibnamefont {Hwang}}, \ and\
  \bibinfo {author} {\bibfnamefont {E.}~\bibnamefont {Rossi}},\ }\href@noop {}
  {\bibfield  {journal} {\bibinfo  {journal} {Rev.~Mod.~Phys.}\ }\textbf
  {\bibinfo {volume} {83}},\ \bibinfo {pages} {407} (\bibinfo {year}
  {2011})}\BibitemShut {NoStop}%
\bibitem [{\citenamefont {Cheng}\ \emph {et~al.}(2010)\citenamefont {Cheng},
  \citenamefont {Song}, \citenamefont {Zhang}, \citenamefont {Zhang},
  \citenamefont {Wang}, \citenamefont {Jia}, \citenamefont {Wang},
  \citenamefont {Wang}, \citenamefont {Zhu}, \citenamefont {Chen},
  \citenamefont {Ma}, \citenamefont {He}, \citenamefont {Wang}, \citenamefont
  {Dai}, \citenamefont {Fang}, \citenamefont {Xie}, \citenamefont {Qi},
  \citenamefont {Liu}, \citenamefont {Zhang},\ and\ \citenamefont
  {Xue}}]{Cheng-et-al-2010}%
  \BibitemOpen
  \bibfield  {author} {\bibinfo {author} {\bibfnamefont {P.}~\bibnamefont
  {Cheng}}, \bibinfo {author} {\bibfnamefont {C.}~\bibnamefont {Song}},
  \bibinfo {author} {\bibfnamefont {T.}~\bibnamefont {Zhang}}, \bibinfo
  {author} {\bibfnamefont {Y.}~\bibnamefont {Zhang}}, \bibinfo {author}
  {\bibfnamefont {Y.}~\bibnamefont {Wang}}, \bibinfo {author} {\bibfnamefont
  {J.-F.}\ \bibnamefont {Jia}}, \bibinfo {author} {\bibfnamefont
  {J.}~\bibnamefont {Wang}}, \bibinfo {author} {\bibfnamefont {Y.}~\bibnamefont
  {Wang}}, \bibinfo {author} {\bibfnamefont {B.-F.}\ \bibnamefont {Zhu}},
  \bibinfo {author} {\bibfnamefont {X.}~\bibnamefont {Chen}}, \bibinfo {author}
  {\bibfnamefont {X.}~\bibnamefont {Ma}}, \bibinfo {author} {\bibfnamefont
  {K.}~\bibnamefont {He}}, \bibinfo {author} {\bibfnamefont {L.}~\bibnamefont
  {Wang}}, \bibinfo {author} {\bibfnamefont {X.}~\bibnamefont {Dai}}, \bibinfo
  {author} {\bibfnamefont {Z.}~\bibnamefont {Fang}}, \bibinfo {author}
  {\bibfnamefont {X.}~\bibnamefont {Xie}}, \bibinfo {author} {\bibfnamefont
  {X.-L.}\ \bibnamefont {Qi}}, \bibinfo {author} {\bibfnamefont {C.-X.}\
  \bibnamefont {Liu}}, \bibinfo {author} {\bibfnamefont {S.-C.}\ \bibnamefont
  {Zhang}}, \ and\ \bibinfo {author} {\bibfnamefont {Q.-K.}\ \bibnamefont
  {Xue}},\ }\href@noop {} {\bibfield  {journal} {\bibinfo  {journal}
  {Phys.~Rev.~Lett.}\ }\textbf {\bibinfo {volume} {105}},\ \bibinfo {pages}
  {076801} (\bibinfo {year} {2010})}\BibitemShut {NoStop}%
  
\bibitem [{\citenamefont {Arciniaga}\ and\ \citenamefont
  {Peterson}()}]{Arciniaga-Peterson-2016}%
  \BibitemOpen
  \bibfield  {author} {\bibinfo {author} {\bibfnamefont {M.}\ \bibnamefont
  {Arciniaga}}\ and\ \bibinfo {author} {\bibfnamefont {M.~R.}\ \bibnamefont
  {Peterson}},\ }\href@noop {} {\ }\Eprint {http://arxiv.org/abs/1602.03937}
  {arXiv:1602.03937} \BibitemShut {NoStop}%

\bibitem [{\citenamefont {Wooten}\ and\ \citenamefont
  {Macek}()}]{Wooten-Macek-2014}%
  \BibitemOpen
  \bibfield  {author} {\bibinfo {author} {\bibfnamefont {R.~E.}\ \bibnamefont
  {Wooten}}\ and\ \bibinfo {author} {\bibfnamefont {J.~H.}\ \bibnamefont
  {Macek}},\ }\href@noop {} {\ }\Eprint {http://arxiv.org/abs/1408.5379}
  {arXiv:1408.5379} \BibitemShut {NoStop}%




\end{thebibliography}
\end{document}